\documentclass[preprint,journal]{vgtc}       % preprint (journal style)

\ifpdf%                                % if we use pdflatex
  \pdfoutput=1\relax                   % create PDFs from pdfLaTeX
  \usepackage{graphicx}                % allow us to embed graphics files
  \DeclareGraphicsExtensions{.pdf,.png,.jpg,.jpeg} % for pdflatex we expect .pdf, .png, or .jpg files
\else%                                 % else we use pure latex
  \usepackage{graphicx}                % allow us to embed graphics files
  \DeclareGraphicsExtensions{.eps}     % for pure latex we expect eps files
\fi%

\graphicspath{{figures/}{pictures/}{images/}{./}} % where to search for the images

\usepackage{microtype}                 % use micro-typography (slightly more compact, better to read)
\PassOptionsToPackage{warn}{textcomp}  % to address font issues with \textrightarrow
\usepackage{textcomp}                  % use better special symbols
\usepackage{mathptmx}                  % use matching math font
\usepackage{amsmath}
\usepackage{amssymb}
\usepackage{times}                     % we use Times as the main font
\usepackage{cite}                      % needed to automatically sort the references
\usepackage{tabu}                      % only used for the table example
\usepackage{booktabs}                  % only used for the table example
\usepackage{xspace}
\usepackage{comment}
\usepackage{multirow}
\usepackage[usenames,dvipsnames]{xcolor}
\usepackage[normalem]{ulem}
\usepackage{comment}
\usepackage{makecell}
\definecolor{mred}{rgb}{.80,.12,.30}
\definecolor{grey}{rgb}{0.5,0.5,0.5}
\definecolor{outlinecolor}{rgb}{0,0.18,0.38}
\definecolor{Purple}{rgb}{.75,0,.85}
\definecolor{orange}{rgb}{.81,.29,.13}
\definecolor{green}{rgb}{.11,.68,.15}

\newif\ifnotes
\notestrue

\newcommand{\oursystem}{NeuralCubes\xspace}

\let\origcite\cite
\renewcommand{\cite}[1]{\ifnotes\mbox{\origcite{#1}}\else \origcite{#1}\fi}

\newcommand{\mb}{\mathbf}

\preprinttext{ }

\vgtccategory{Research}
\vgtcpapertype{Algorithm/Technique}

\title{ \oursystem: Deep Representations for Visual Data Exploration}

\author{Zhe Wang, Dylan Cashman, Mingwei Li, Jixian Li, Matthew Berger, \\
Joshua A. Levine, Remco Chang, and Carlos Scheidegger}

\authorfooter{
\item Zhe Wang, Mingwei Li, Jixian Li, Josh Levine and Carlos Scheidegger are with the University of Arizona. emails: \{zhew, jixianli, mwli, josh, cscheid\}@email.arizona.edu.
\item Matthew Berger is with Vanderbilt University. email: matthew.berger@vanderbilt.edu. 
\item Dylan Cashman and Remco Chang are with Tufts University. email: \{dylancashman, remco\}@tufts.edu.
}

\shortauthortitle{Wang et al., \oursystem: Deep Representations for Visual Data Exploration}

\abstract{
	Visual exploration of large multidimensional datasets has seen tremendous progress in recent years, allowing users to express rich data queries that
	produce informative visual summaries, all in real time.
  Techniques based on data cubes are some of the most promising approaches.
  However, these techniques usually require a large memory footprint for large datasets.
  To tackle this problem,
  we present \oursystem: neural networks that predict results for aggregate queries, similar to data cubes.
	\oursystem learns a function that takes as input a given query, for instance, a geographic region and temporal interval, and outputs the result of the query.
	The learned function serves as a real-time, low-memory approximator for aggregation queries.
  \oursystem models are small enough to be sent to the client side (e.g. the web browser for a web-based application) for evaluation, 
  enabling data exploration of large datasets without database/network connection.
  We demonstrate the effectiveness of \oursystem through extensive experiments on a variety of datasets 
  and discuss how \oursystem opens up opportunities for new types of visualization and interaction.
} % end of abstract


\CCScatlist{ % not used in journal version
 \CCScat{K.6.1}{Management of Computing and Information Systems}%
{Project and People Management}{Life Cycle};
 \CCScat{K.7.m}{The Computing Profession}{Miscellaneous}{Ethics}
}

\teaser{
  \centering
  \includegraphics[width=0.9\linewidth]{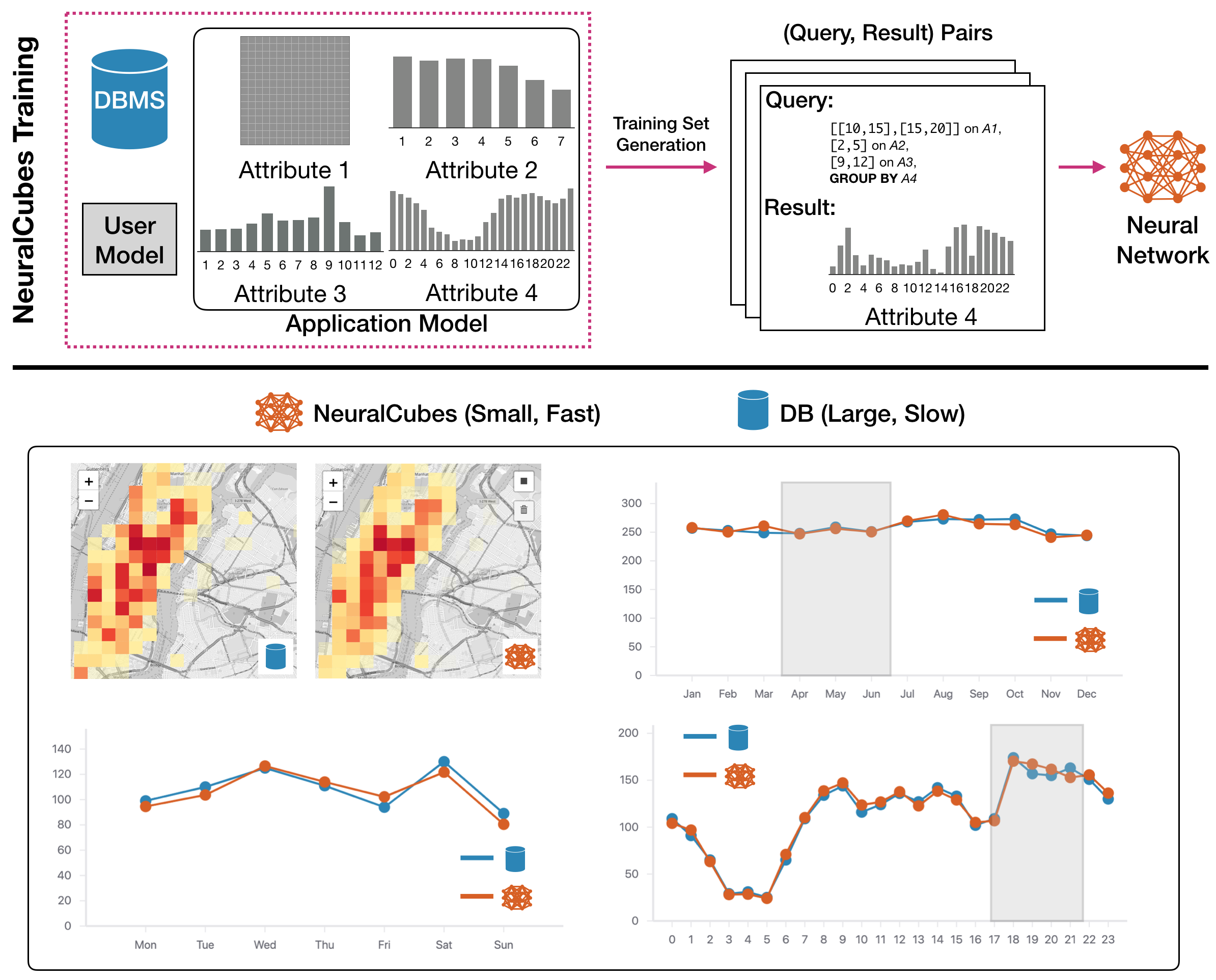}
  \caption{
    \oursystem is a system that removes the need for a visualization to connect to a database during runtime.
    Instead, \oursystem approximate the query results of a database with a trained neural network model.
    (Top): At the offline training stage, \oursystem use a specific user model (types and frequency of different queries they will make) and application model to generate query-result pairs.
    Then a neural network model is trained on these pairs so that it learns how to predict a result for a future new query.
    At runtime, the trained model answers all queries issued by the application.
    (Bottom): The example shown in the lower part of the figure is a data visualization system for YellowCab taxi data\cite{yellowcabdata}.
    The orange lines and the heatmap on the right are plotted using predictions by the trained model.
    For the purpose of evaluation, we also plot the ground truth by blue lines and by the heatmap on the left.
    Note that the query results generated from NeuralCubes closely match the ground truth.
    However, trained NeuralCubes is small in memory footprint.
    In the example above, the NeuralCubes requires only 798KB whereas the raw YellowCab taxi dataset used in the example is 1.8GB.
  }
    
  \label{fig:teaser}
}

\vgtcinsertpkg

\begin{document}


\firstsection{Introduction}

\maketitle

% visual data exploration
Interactive visual exploration is becoming increasingly essential for making sense of large multidimensional datasets.
%% This is especially true of modern datasets, where the size and dimensionality of data is rapidly increasing.
It is not uncommon for datasets to have billions of data items that contain a variety of attributes of geographic, temporal,
and categorical nature.
Due to its size and complexity, querying such large data in real-time is often not feasible as it will result in unreasonable amount of latency.
Instead, efficient data structures are used by visualizations in lieu of querying raw data in databases in real-time. These data structures are pre-computed and optimized around queries that are frequently used by the visualization, such as performing summary (aggregation) of the data\cite{lins2013nanocubes, pahins2017hashedcubes}, ranking\cite{miranda2018topkube}, and applying multivariate statistics\cite{wang2017gaussian}.

However, while these data structures are effective, they can still be prohibitively large as data size increases. 
Worse, when data complexity increases (i.e. in terms of the number of dimensions in the data), the sizes of many of these data structures grow exponentially. 
As a result, these data structures are often stored on a server. Only the sub-parts of the data structures are fetched in real-time based on the user’s exploration.

%% observation and technique
%It is useful to think of existing techniques as \emph{processes} that map from a given input query to produce an aggregation
%result.
%We propose to use deep neural networks as a \emph{model for answering aggregation queries}.
%Our approach, which we call \oursystem, learns this process by optimizing a neural network to predict aggregations by training on a large collection of aggregate query results.
%The result of training \oursystem yields a surrogate model for performing aggregation queries, which runs in real time and consumes a small amount of memory
%--- dependent of the number of parameters of the neural network rather than the database schema or input dataset --- 
%and enables the user to visually explore data in much the same way as one would use existing systems.

In this paper, we introduce NeuralCubes, a technique that generates extremely small “data structures” that can support interactively exploration of multi-dimensional data. 
Unlike existing data structures that rely on performing and storing summary statistics, NeuralCubes is a trained deep neural network that can respond to queries about the data in real time. 
We design a compact neural network architecture yet accurate enough for data visualization.
Due to its extremely small footprint, NeuralCubes can be stored in client memory, 
thereby eliminating the need for a visualization system to fetch data or sub-parts of a data structure from a server in real-time. 
Instead, with NeuralCubes, all queries can be computed on the client in real time.

% application
%At runtime, the learned model can be sent to the client side, e.g. a web browser, for evaluation,
%eliminating the need for a backend server.
%Without network dependency, \oursystem can provide a more responsive data exploration user experience. 
%In addition, since we have control over the \emph{architecture} of the neural networks in \oursystem, we can design the architecture such that it encodes interesting structures of the dataset and then expose what the network learns to the user for exploration.
%For example, \oursystem can learn 2D projections for each of the inputs that are \emph{predictive} of the aggregation query results.
%We expose this to the user as a set of linked scatterplot views, where the user can explore the \emph{space of queries}, driven by the similarity of queries
%in their aggregations.

The inspiration behind our design of NeuralCubes is that we consider querying a database as a function that maps from a given input query to produce an aggregation result. 
Assuming that there are latent patterns in the data (i.e. that the data is not purely random), 
these functions can be efficiently learned using the latest advances in deep learning. 
In particular, we observe that typical visualizations (such as the one shown in Fig.~\ref{fig:teaser}) generate a limited number of query templates and expect a fixed number of numeric values in response. 
For example, in the example shown in Fig.~\ref{fig:teaser}, the query to the database will be based on four sets of filters (geographic region, month of year, day of week, and time of day). 
In response, the visualization anticipates a set of numeric values to populate the geographic heatmap and the three line charts.
Given this clearly defined inputs and outputs of the function, deep neural networks have been shown to be robust and efficient in learning the mapping.

%To train a NeuralCube, blah blah blah (Remco: we need one more paragraph of technical depth in how NeuralCubes are built, and how they are used in run time)
\oursystem has a off-line training stage and a real-time running stage.
To train \oursystem, we first need an application model and a user model to generate training set.
An application model can be seen as the ``data schema'' of an application.
It contains information of how many attributes are used and what's the format of each attribute.
A user model derives from types of queries and the frequency of them that users perform when using an application.
With these two models, we can easy generate query-result pairs as training set for \oursystem.
We use many-hot encoding to represent the input query and feed it to neural networks that try to predict the result of that query.
After training, the learned neural network model can answer any queries issued from the same application.
Since the model is very small, \oursystem model can be evaluated in real-time on any modern CPU/GPU.

%(Remco: need a section on evaluation) We evaluate NeuralCubes blah blah blah
We evaluate \oursystem on a variety of datasets including BrightKite social network check-ins\cite{cho2011friendship}, 
Flights dataset\cite{flightdata}, YellowCab taxi dataset\cite{yellowcabdata}, and SPLOM dataset\cite{kandel2012profiler}.
We quantitatively analyze the accuracy of \oursystem and how neural network size, training set size, raw data size, and attribute resolution affect prediction.
We report experimental results in Section~\ref{sec:case_studies}.

% bullets
We summarize our contributions as follows:
\begin{itemize}
\item we show that neural networks can learn to answer aggregate queries efficiently and effectively, that they generalize across heterogeneous attribute types (such as geographic, temporal, and categorical data), and present a method to convert the schemata needed to describe visual exploration systems into an appropriate deep neural network architecture;
\item we use these neural networks that learn the structure of aggregation queries to provide the user 2D projections that enable the intuitive exploration of data queries; and
\item we conduct extensive experiments on a variety of datasets that proves the effectiveness of our approach.
\end{itemize}

\section{Related Work}

Our work proceeds from recent work in two mostly disparate fields; data management and neural networks.
In data management, we discuss architectures, data structures, and algorithms that exploit access patterns to offer better performance.
In neural networks, we review some of the recent applications of neural networks to novel domains, as well as relevant work on the interpretability of deep networks.

\subsection{Data management}

The importance of data management technology in the context of interactive data exploration has been recognized for over 30 years, with the work of MacKinlay, Stolte, and collaborators in Polaris, APT, and Show Me being central contributions to the field\cite{mackinlay1986automating, stolte2002polaris, mackinlay2007show}. Since then, researchers in both data management and visualization have extended the capabilities of data exploration systems (both visual and otherwise) in a number of ways. Scalable, low-latency systems now exist for both the exact and approximate querying situation\cite{zaharia2010spark, agarwal2013blinkdb}. Recently, Wu et al. have proposed that data management systems should be designed specifically with visualization in mind\cite{wu2014case}. \oursystem, as we will later discuss, provides evidence that machine learning techniques should \emph{also} be designed with visualization in mind, and that such design enables novel visual data exploration tools.

While we developed \oursystem\ to leveraging machine learning technology for providing richer information during data exploration itself, we are clearly not the first to propose to use machine learning techniques in the context of data management. Notably, ML has been recently used to enable \emph{predictive interaction}: if a system can accurately predict the future behavior of the user, there are ample opportunities for performance gains (and specifically for hiding latency)\cite{chan2008maintaining, battle2016dynamic}.

Gray et al.'s breakthrough idea of organizing aggregation queries in the appropriate lattice --- the now-ubiquitous \emph{data cube} --- spawned an entire subfield of advances in algorithms and data structures\cite{gray1997data, harinarayan1996implementing, sismanis2002dwarf}. This work has gained renewed interest in the context of interactive exploration, where additional information (such as screen resolution, visualization encoding, and query prediction) can be leveraged\cite{liu2013immens, lins2013nanocubes, kamat2014distributed, pahins2017hashedcubes}. Since every query in \oursystem is executed by a \emph{fixed-size network}, it also provides low latency in aggregation queries. But because we design \oursystem specifically so that networks learn the interaction between query inputs and results, it provides additional information about the dataset that can itself be used in interactive exploration.

\subsection{Deep Neural Networks}

% deep learning in general
Our approach is inspired by the recent success of applying deep neural networks to a variety of domains, including image recognition\cite{NIPS2012_4824}, machine
translation\cite{sutskever2014sequence}, and speech recognition\cite{hinton2012deep}. These techniques are solely focused on prediction, and our method is similar,
in that we are focused on training deep networks for the purposes of query prediction.
Yet, we differ in that prediction is not the only goal, 
rather we want to perform learning in a manner that provides the user a fast and low-memory-cost way to visually explore data.
The query prediction task at hand can be viewed as a means to realize these goals.

% VA for interpretability in deep learning
Our approach to training neural networks for data exploration can be viewed as a form of making deep networks more interpretable.
In the visual analytics (VA) literature there has been much recent work devoted to the interpretability of deep networks, namely with respect to
the training process, the learned features of the network, the input domain, and the network's output space.
Liu et al.\cite{liu2017towards} visualize convolutional neural networks (CNNs) by clustering activations within each layer given a set of inputs,
and visualize prominent flows between clustered activations.
Other VA approaches to interpreting CNNs have considered visualizing correlations between classes during the training process\cite{bilal2018convolutional},
and visualizing per-layer convolution filters and their relationships, in order to understand filters that are important to training\cite{pezzotti2018deepeyes}.
Visualizing and understanding recurrent neural networks (RNNs) has also received much attention, through understanding the training process\cite{cashman2017rnnbow},
as well as understanding hidden states dynamics\cite{strobelt2018lstmvis} .
All of these approaches seek to provide interpretability for deep neural networks that were never designed to be interpretable.
In contrast, our approach \emph{directly builds interpretability} into the network, such that the user can take advantage of different aspects
of the learned network to help their exploration.

\begin{figure}
	\centering
	\includegraphics[width=1\linewidth]{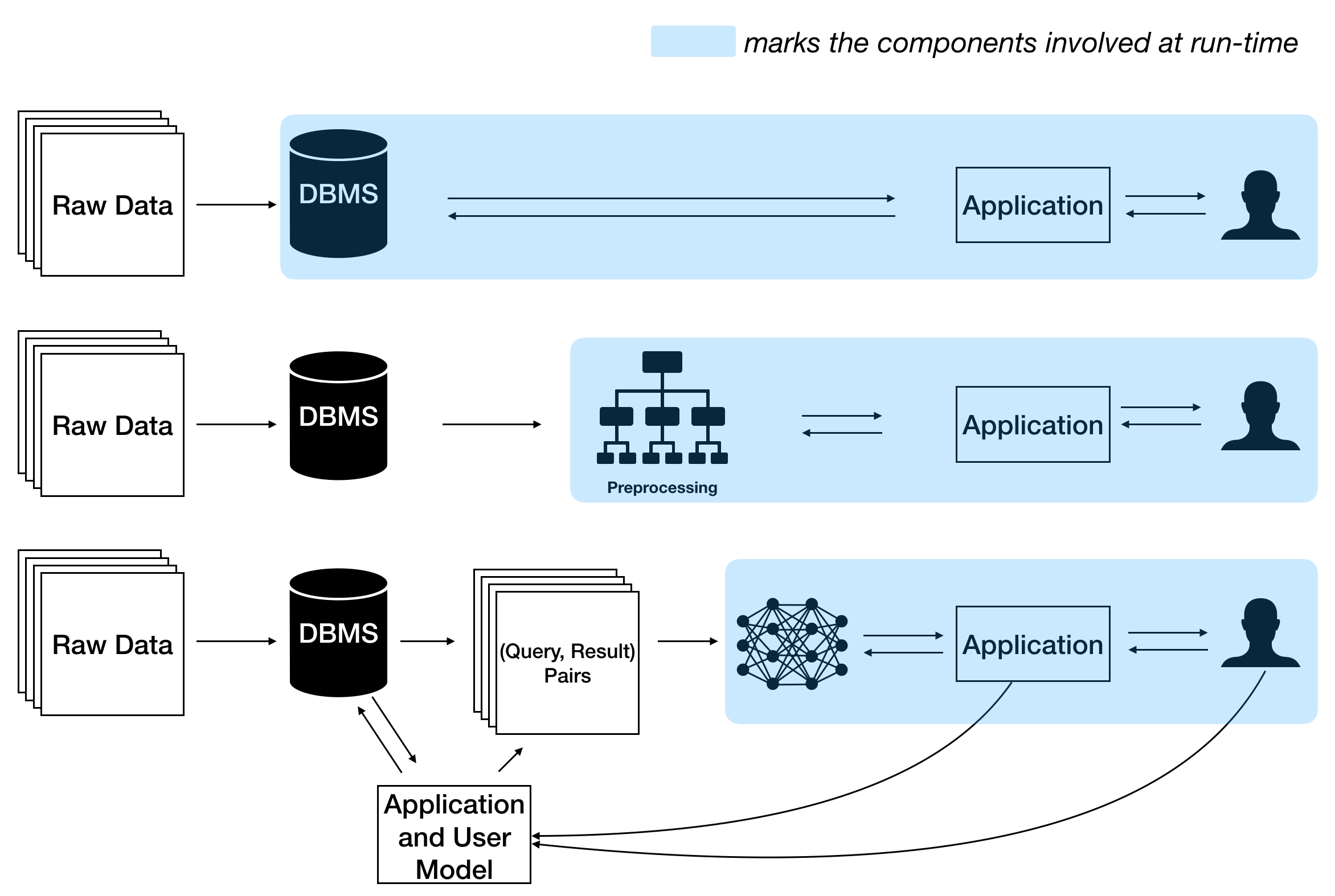}
  \caption{Comparison of systems using traditional database, Datacubes and \oursystem.}
	\label{fig:sys_comparison}
  \vspace{-1em}
\end{figure}

% self supervision
In this context, our method for learning features of aggregation queries can be viewed as a form of unsupervised learning, where treating query prediction as pretext,
the features that we learn along the way can be used for other purposes -- in our case exploratory purposes.
This is similar to recent techniques in computer vision that learn features using different forms of self supervision, for instance learning to
predict spatial context\cite{Doersch_2015_ICCV,noroozi2016unsupervised}, temporal context\cite{Wang_2015_ICCV}, and perhaps more pertinent to our work, learning to count
visual primitives in a scene\cite{Noroozi_2017_ICCV}.
These techniques solve certain types of relevant visual tasks that do not require human supervision, but then extract the learned features for supervised learning.
Our approach is similar: our training data does not require human intervention, since it is built from existing data cubes techniques, yet the features that
we learn from this task can be used to help with visual data exploration.

We also note that there is some very recent work that seeks to combine databases with neural networks.
Kraska et al.\cite{kraska2017case} make the connection between indexing, such as b-trees or hashes, and models, and show that such indexing schemes
can be learned using neural networks.
Mitzenmacher\cite{mitzenmacher2018model} consider similar learning techniques for Bloom filters.
These methods are concerned with using neural networks to speed up computation and minimize memory storage.
Although we demonstrate that our method can attain these benefits, 
the primary focus of our method is in using a neural network
as an integral component to visual exploration, 
i.e. \oursystem is not trying to predict \textit{any} queries that a database can answer.

\section{\oursystem: Replacing a Database with a Learned Neural Network}

We introduce our approach by comparing it with visualization systems that utilizing traditional database and those using advanced preprocessing techniques. 
Fig.~\ref{fig:sys_comparison} is an overview of these different approaches.

% traditional databases
First, we briefly discuss what data queries does a interactive visualization system need.
Suppose we are given a set of \emph{records}, each record contains a set of \emph{attributes}, and each attribute has a certain \emph{type},
for instance continuous, categorical, geographic, and temporal, that characterizes the set of \emph{values} it may take on.
Database queries may return a single record, or multiple records, and in the case of the latter it is often of interest to summarize the set of records
by performing an \emph{aggregation}, for instance \textsf{count}, \textsf{average}, or \textsf{max}, depending on the attribute type.
Within a visualization system, the set of attributes, their types, and the class of aggregations determine the sorts of queries one may issue
that serve as the backbone for visual interaction.
For instance, we may perform a \textsf{group-by} query for a given attribute that will return, for each of its values, the result of a specified aggregation, e.g. \textsf{count}.
If the attribute type is categorical or temporal, then we can visualize this result as a histogram, whereas if the attribute type is geographic, 
we may plot the result as a heatmap over a spatial region.

\begin{figure}
	\centering
	\includegraphics[width=1\linewidth]{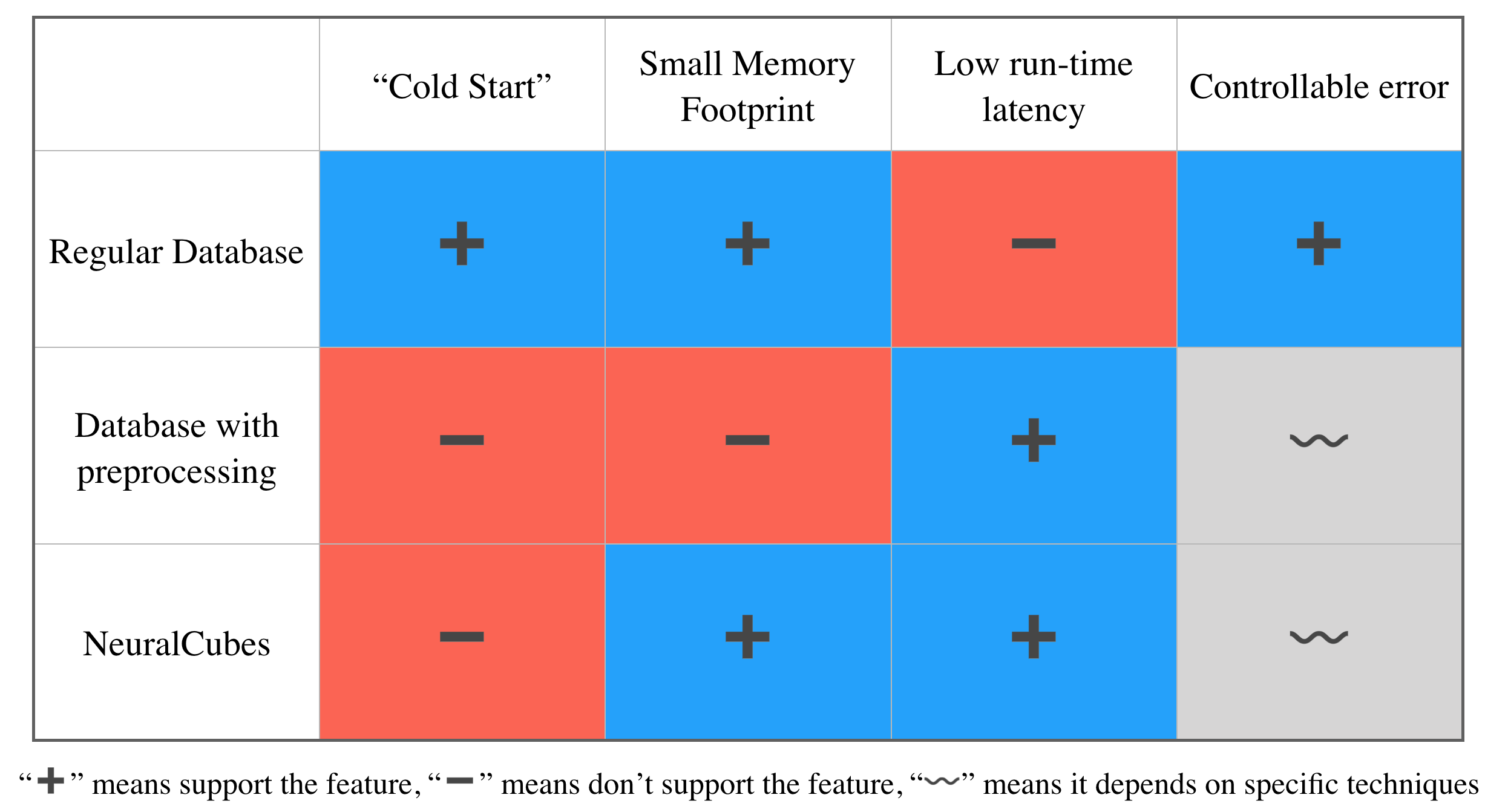}
  \caption{Comparison of supported features of different approaches.}
	\label{fig:feature_comparison_table}
  \vspace{-1em}
\end{figure}

Naively, a visualization system can always issue SQL queries to the database to get the needed information.
As summarized in Fig.~\ref{fig:feature_comparison_table}, this approach does support ``cold start", which
means no extra processing is needed when the data changes (e.g. adding new records).
Besides, there is no need for extra memory other than the data set itself.
The results will always be accurate.
However, this approach obviously won't scale when the dataset gets larger,
which stops it from being practically used in interactive visualization systems for large dataset.

% databases for visualization
Many database techniques have been proposed with visualizations in mind.
For example, datacubes based proposals\cite{lins2013nanocubes,pahins2017hashedcubes,wang2017gaussian} are built with the goal of answering aggregational queries in real-time. 
In particular, they are focused on linked views, such that selection of an attribute in one view updates the visualization of other views, where view updates
are based on a set of queries made to the database.
Linked views enable the user to engage with the visualization in an iterative process: the user spots a trend in one view, makes a selection based on the trend
(i.e. filter on a geographic region), they are presented with updated views in the remaining attributes, and their interaction process repeats.
However, these techniques usually requires a large memory footprint and long preprocessing time.
While some datacubes based technique can provide absolute correct answer at a given resolution,
many sampling based techniques can not guarantee controllable error.

% \oursystem
The basic idea behind \oursystem is the use of neural networks to learn the process of performing database aggregation queries.
It is thus useful to think about the neural network as a function that approximates a database aggregation query, 
where the input is a data query in the form of a set of attribute ranges, and the output is the aggregation of the data returned from the given input query.
The fundamental difference of \oursystem with existing techniques is that \oursystem is trying to solve the problem in a pure machine learning perspective.
Thus \oursystem has two modes: offline training and real-time querying.
During offline training, we generate query-result pairs and train neural networks to learn to predict results for queries.
Then at real-time querying stage, new queries will be evaluated by the trained model to get a predicted result. 

% why should this work
%It is natural to ask the following: \textbf{why should we expect neural networks to be able to learn how to query?}
%In particular, what properties of aggregation queries make this problem tractable for ML techniques?
%As an example, consider the \textsf{count} aggregation, which is the total number of results returned from a given query.
%Suppose that we have two queries $Q_i$ and $Q_j$, and without loss of generality suppose that $r_i(a_k) = r_j(a_k)$ for $k > 1$,
%and $r_i(a_1) \cap r_j(a_2) = \emptyset$, i.e. the queries are identical for all attributes except the first, where the attribute ranges are disjoint.
%Then the \textsf{count} aggregation respects the following property: $DB(Q_i \cup Q_j) = DB(Q_i) + DB(Q_j)$.
%This implies that the \textsf{count} aggregation is \emph{multilinear}. In other words, fixing all but one attribute, the union of
%two queries that are disjoint in the held out attribute gives the same result as the addition of the individual queries.
%More generally, replacing \textsf{count} with any other aggregation operation that corresponds to a commutative monoid $\oplus$ will respect
%a generalized version of this property.
%As a result, as long as the aggregation operation itself is relatively well-behaved, we can expect the network to be able to learn the structure of the data aggregation queries.

Like most machine learning techniques, the predicted query results from \oursystem don't have a strict error bound.
However, we can practically control the error by using neural networks with enough capacity and feeding in a large enough training set.
Furthermore, \oursystem is designed for visualization.
The absolute error is visually neglectable as long as the overall trend and distribution reflects the truth.
User can always issue SQL queries to the database to get the accurate numbers.
%More importantly, the trained model is very small.
%It can be sent to the client side for evaluation.
%This totally eliminates network overhead, from which traditional techniques suffers a lot.
%The value of \oursystem lies in that it provides a ``light'' yet powerful tool for users to explore the dataset before they know exactly what they want to find.

We summarize the trade-offs of different approaches discussed above in Fig.~\ref{fig:feature_comparison_table}.

\subsection{What's the input of \oursystem?}
\label{sec:input_of_model}

\begin{figure}[t]
	\centering
	\includegraphics[width=0.9\linewidth]{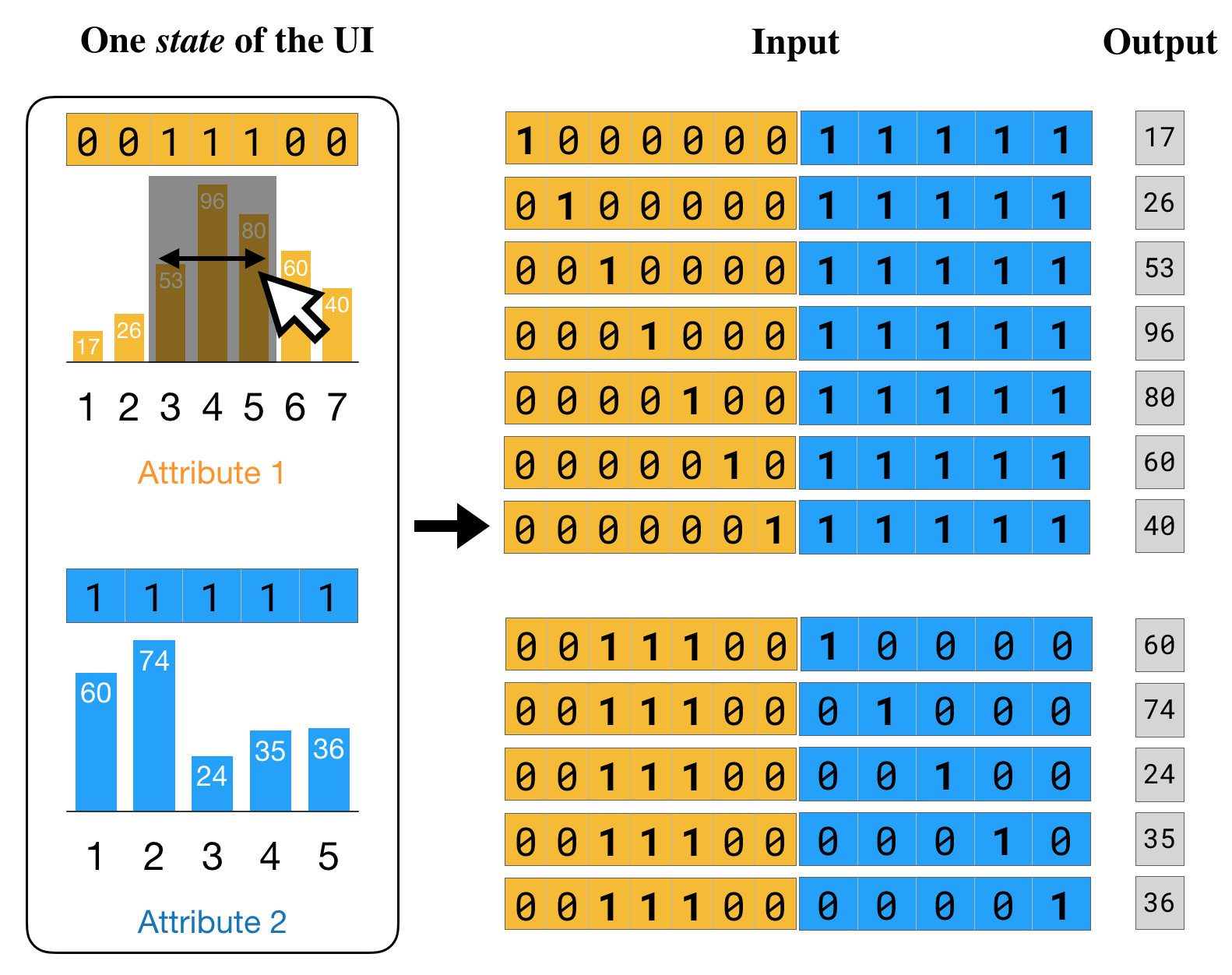}
  \caption{The corresponding input-out pairs for one \emph{state} of the UI.}
	\label{fig:input_encoding}
  \vspace{-1em}
\end{figure}

% overview - notation, database aggregation queries as functions, objective is to: regress on aggregations, learn query features/projections

We first define the concept of \textit{state} of a data visualization system.
We assume that the underlying database schema has a total of $d$ attributes, where we denote each attribute by $a_i , 1 \le i \le d$, and we represent the \emph{range selection} operation for a given attribute
$a_d$ by $r(a_d)$.
For instance, if an attribute was hour-of-day, then the range operation on this attribute would return a set of hours.
At any time, there must be a range selected on each attribute. 
(No selection on a attribute means select the full range of this attribute.)
The set of these ranges of different attributes determine what the plots of each attribute look like.
We call this set of ranges a \textit{state} of the visualization system,
denoted as $S = {r(a_1), r(a_2), ... r(a_d)}$.
The corresponding query results are $DB(S) \in \mathbb{R}$.
%Given this notation, we denote a database query for a specific type of aggregation as $DB(s(a_1), s(a_2), ... s(a_d)) \in \mathbb{R}$.
Our objective is to train a neural network $f(S) \in \mathbb{R}$ to best approximate $DB$, given \emph{training data}
$D = \left((S_1,DB(S_1)), (S_2,DB(S_2)), ... (S_n,DB(S_n))\right)$, 
where $S_i$ is a state, i.e. $S_i = \left(r_i(a_1), r_i(a_2), ... r_i(a_d)\right)$,
and $DB(S_n)$ is the aggregational result from the database.
For example, the state shown in Fig.~\ref{fig:input_encoding} is $S=\{[3,5], [1,5]\}$.

%At any time, there must be a range selected on each attribute. 
%(No selection on a attribute means select the full range of this attribute.)
%The set of these ranges of different attributes determine the what the plots of each attribute look like.
%We call this set of ranges a \textit{state} of the visualization system.
%Apparently, each \textit{state} corresponds to a specific set of queries.
%The intuition behind this strategy is that we want the neural network to see the whole picture of system rather than several
%independent queries.

%\subsubsection{Query Input Representation}

% how do we represent selections? many-hot encodings -> extending one-hot encodings (refs: neural language models, encoding spatial queries in visual recognition)
Then we determine the representation of the range selection that is fed into the network.
This is nontrivial due to the different types of attributes, e.g. geographic, temporal, categorical, as well as the types of selections that can be performed
on attributes, e.g. spatially contiguous selections in geographic coordinates.
To address these challenges in a unified manner, we use \emph{many-hot encodings} for attribute selections, as shown in Fig.~\ref{fig:input_encoding}.
Many-hot encodings are generalizations of one-hot encodings, commonly used as a way to uniquely represent words in neural language models\cite{bengio2003neural},
categorical inputs for generative models\cite{dosovitskiy2015learning}, as well as geographic coordinates for image recognition\cite{tang2015improving}.

% details
More specifically, for a given attribute $a_i$ we assume that it may be discretized into $m(a_i)$ many values.
For certain attributes, this assumption is natural: categorical data, temporal data such as hour-of-day or day-of-week, while for
continuously-valued data we uniformly discretize the data space into bins.
For the attribute's selection $r(a_i)$, we then associate a binary vector $\mb{r}(a_i) \in \{0,1\}^{m(a_i)}$ such that $\mb{r}(a_i)_j = 1$ if the value
at index $j$ belongs to the selection, and $0$ otherwise.
This permits arbitrary types of selections for categorical and temporal data.
Spatial data, specifically 2D geographic regions, is slightly more complicated: one option is to represent each discretized cell
as a single dimension in $\mb{r}$, but this would result in a large number of inputs for even small spatial resolutions.
We simplify this problem by restrict selections on 2D regions to be rectangular.
Thus to represent such a region we associate a pair of vectors $\mb{r}^x(a_i)$
and $\mb{r}^y(a_i)$ to for the selected x and y intervals, respectively, of the rectangle and then concatenate these two vectors to form the input.
In practice, non-rectangular selections can be approximated by issuing multiple rectangular selections. 

When generating training set, the ground truth results for the queries are obtained from making actual database queries.
At runtime, new queries will be encoded in the same way as in training stage.
Then it will be fed into the trained model to get a predicted result.

\begin{figure}[t]
	\centering
  \includegraphics[width=1.0\linewidth]{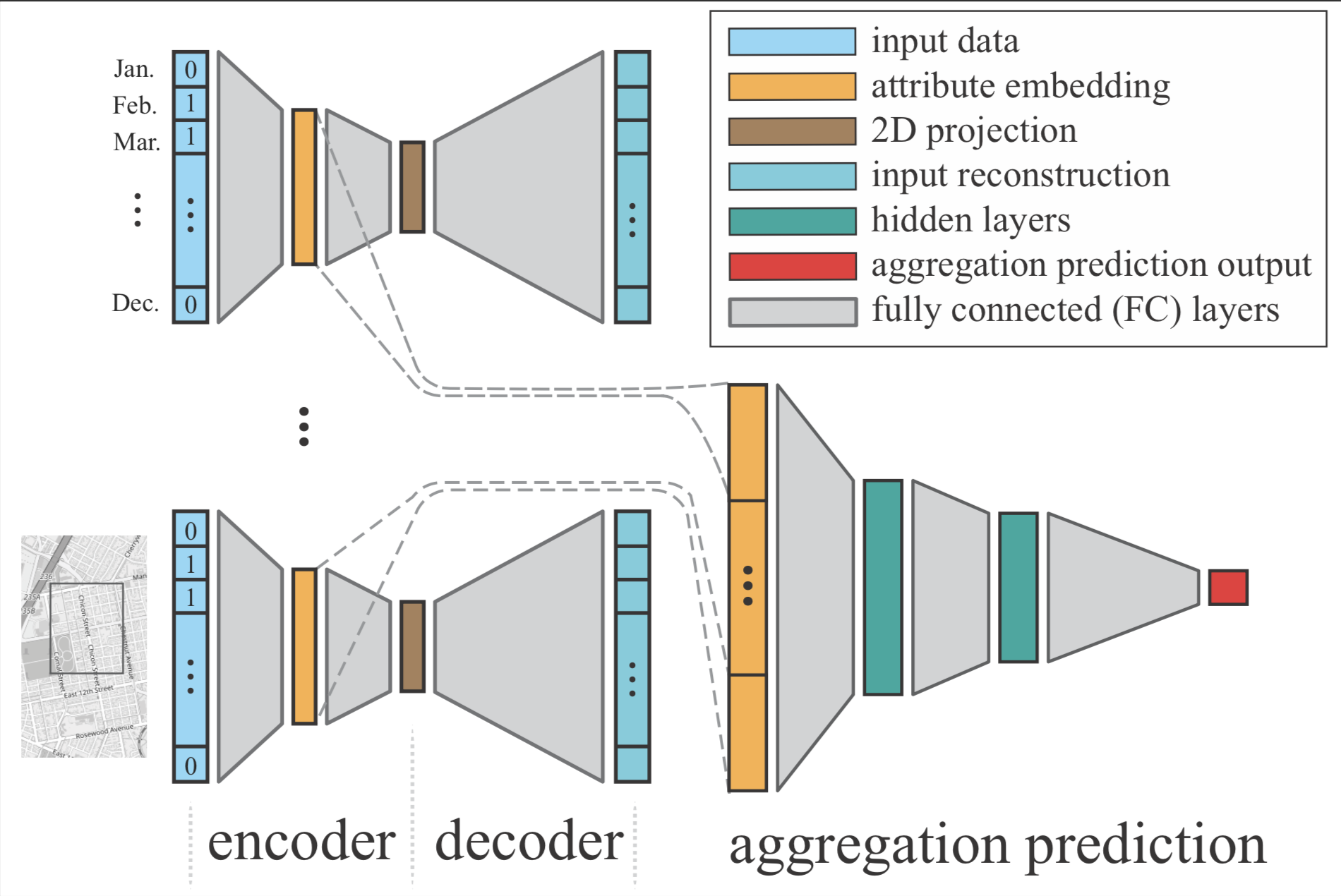}
	\caption{We highlight the general structure of our neural network. For each attribute, we first learn a feature embedding (orange), and
	then use the embedding for two purposes: we concatenate the embeddings to predict aggregations (red), as well as learn a 2D projection (brown).}
  \label{fig:nn_architecture}
  \vspace{-1em}
\end{figure}

\subsection{Generating Training Data: Modeling Application and User Interaction}

\subsubsection{Collecting vs. Generating Training Set}

Machine learning models can be seen as a function of the data that they are trained on\cite{koh2017understanding}.
It is of integral importance that the training set reflects the goal of the network.
In many applications of neural networks such as image classifiers, training data sets must be gathered from the real world,
and manually labeled by a large number of human workers\cite{deng2009imagenet}.
In contrast, since \oursystem are used to approximate database queries, a training set can instead be generated by
executing queries against a dataset to form ground truth.  

\subsubsection{Sampling in Input Space vs. User Query Space}

When generating a training set, it is important to be careful \textit{how} the queries are generated.
The space of potential queries is exponential in the size of the range of values that can be queried over.
But within an information visualization, some queries are much more likely, and thus much more important for the
neural network to predict accurately.
By choosing a sampling strategy that mirrors the types of queries that will be called by the visualization, we can
focus the learning problem on the relevant data distribution.

%Since most existing data visualization applications tend to visualize aggregation results via histograms and heatmaps~\cite{lins2013nanocubes,liu2013immens},
%we would like our sampling strategy to reflect these use cases.
%In this case, queries are uniquely determined by the attributes chosen by the user, and resulting selections over those attributes.
As discussed in section~\ref{sec:input_of_model}, we should randomly generate a \textit{state} of the UI and then turn it into corresponding queries.
Thus, to generate queries, we first generate a range selection for each attribute, e.g.
contiguous ranges for temporal or spatial attributes.
Then we perform a \textsf{group by} query on one of the attribute with the constrains (range selections) on other attributes, 
resulting in a batch of query-result pairs for the current attribute.
We do the same thing for every attribute,
thus giving us all query-result pairs of a \textit{state}.

In order to sample a range selection over an attribute, there are two strategies we can apply:

\begin{enumerate}
  \item We can uniformly sample a lower bound of the range from all possible values of the attribute and then uniformly choose a valid upper bound.
For example, to generate a range selection for \textit{month} (represented by integers from $1$ to $12$), we randomly choose a lower bound, say $9$.
Then the all possible (exclusive) upper bounds are $10$, $11$, $12$ and $13$.
So we just randomly choose one from them, say $11$. Finally this range selection will be $[9, 11)$.

  \item We can uniformly sample the length of the range from all valid lengths and then uniformly choose a start and end position for that range.
    Using \textit{month} as an example again. The possible length of ranges we can make for \textit{month} are from $1$ to $12$.
    We randomly choose a length, say $3$. Then, for a length-three range, the possible inclusive starting points/lower bounds are $1,2,3,4,5,6,7,8,9,10$. 
    So we randomly choose one from them, say $2$. Finally this range selection will be $[2, 5)$.

\end{enumerate}

The difference between these two strategies is that strategy 1 generates more short-length ranges while strategy 2 generate more long-length ranges.
In our experiments, we found people are likely to make queries that has more long-length ranges.
An extreme example is the most frequently issued query - the default view for a data visualization dashboard, for which we select full lengths at every attribute.
So we apply strategy 2 for all the use cases shown in Section ~\ref{sec:case_studies}.

While it may seem artificial to carefully sample a training set to make the network fit a certain kind of input,
it's important to remember that \oursystem is designed, first and foremost, with visualization in mind.
Thus, even if we aren't necessarily learning over the full data distribution of queries, so long as our sampling resembles the manner in which users perform selection,
then a user's interaction with the network should remain meaningful.

\subsection{A Neural Network Architecture for Data Queries}

% general structure of the neural network
Our neural network is composed of a sequence of \emph{layers}, where a layer is defined as the application of an affine function, followed by applying an elementwise nonlinear function.
In this paper we exclusively use \emph{fully connected} layers.
A fully connected layer at index $i$ is parameterized by a weight matrix $\mb{W}_i \in \mathbb{R}^{b \times a}$ and bias vector $\mb{b}_i \in \mathbb{R}^{b}$,
where it is assumed the previous layer is a vector of length $a$, and the output of the layer produces a vector of length $b$.
We denote the affine function at layer $i$ by $g_i : \mathbb{R}^a \rightarrow \mathbb{R}^b$, the nonlinearity by $h_i : \mathbb{R}^b \rightarrow \mathbb{R}^b$, and thus
the function $f$ is: $f = h_n \circ g_n \circ h_{n-1} \circ g_{n-1} \dots h_1 \circ g_1$.
Our objective then is to find the sets of parameters, namely the weight matrices and biases, for each layer that result in $f$ being a good approximator of $DB$.

%In building a neural network for $f$, there are three key issues to address:
%\begin{enumerate}
	%\item How do we represent a query input?
	%\item How do we design and train the network to predict an aggregation query?
	%\item How do we design and train the network to learn both a feature space, and its 2D projection, of attribute queries?
%\end{enumerate}
%We address each in the following.

% other ways to represent inputs: intervals are bad (optional)

\subsubsection{Predicting Aggregations}

% transform each attribute individually, yields an embedding for each attribute -> merge embeddings, transform until we get to a single value
Our general strategy for predicting aggregations is to learn an embedding for each type of attribute selection, followed by concatenating
the attribute embeddings, and then predicting the aggregation query from the concatenation.
The intuition behind this architecture is to first learn attribute-specific features that are \emph{predictive of aggregation queries}, providing
us a more informative representation than the input many-hot encodings, and then to combine these features to learn their relationships in predicting the aggregation.
We use this general architecture for all of the datasets in the paper, shown in Fig.~\ref{fig:nn_architecture}, but tailor the architectures based on
the given dataset, which we defer to Section 6.
All networks, nonetheless, share the following steps to form the network $f$:
\begin{enumerate}
	\item \textbf{Learning Attribute Embeddings.} For a given set of attribute selections represented as binary vectors, we first transform each of them separately
		into their own feature embedding. Namely, for attribute $a_i$, let $f_i : \mathbb{R}^{m(a_i)} \rightarrow \mathbb{R}^{d_i}$ represent a series of layers
		that transforms the attribute selection to a $d_i$-dimensional embedding space.
	\item \textbf{Attribute Embedding Concatenation.} We then concatenate the embeddings into a single vector $\mb{\hat{f}} = [ f_1(\mb{r}(a_1)) , f_2(\mb{r}(a_2)) , ... f_d(\mb{r}(a_d)) ]$,
		where $\mb{\hat{f}} \in \mathbb{R}^{\hat{d}}$, $\hat{d} = \sum_{i=1}^{d} d_i$.
	\item \textbf{Aggregation Query Prediction.} Given the concatenated embedding $\mb{\hat{f}}$, we then feed it through a series of layers, where the last layer outputs
		a single value, corresponding to the aggregation query. Multiple fully connected layers are used in order to learn the relationship between the attributes, so as to
		make better predictions.
\end{enumerate}

% loss: combination of L1 and L2 losses
\textbf{Prediction Loss.} Given the neural network $f$, we can now optimize over its set of parameters to best predict database queries $DB$.
For this purpose, we define a loss function for prediction that combines an L1 loss and a mean-squared loss for a given query $Q$:
\begin{equation}
	L_{pred} = \lambda_1 | f(Q) - DB(Q) | + \lambda_2 (f(Q) - DB(Q))^2,
\end{equation}
where $\lambda_1$ and $\lambda_2$ weight the contributions of the L1 and mean-squared losses, respectively.
The intuition behind this loss is to learn the general trend in the data, captured by the L2 loss, but in order for the training to not be overwhelmed by
aggregations that result in very large values, the L1 loss provides a form of robustness.

\subsubsection{Autoencoder: Reconstruction as Regularization}
\label{subsubsec:proj}

% motivation
%Thus far, we have a method for predicting query aggregations.
%However, given that visual exploration is our primary downstream task, a more careful consideration of how to use the network for visualization can better serve our purpose.
%To this end, we would like to \emph{learn 2D projections} for each of the attributes, and use these learned projections for exploring
%the space of queries in a visual interface (c.f. Section~\ref{sec:nnvis}).
Deep neural networks are easy to overfit because they have large amount of parameters.
To avoid that, we need add some regularization in training.
Reconstruction as regularization has been used in many recent work\cite{kingma2014semi, sabour2017dynamic}.
We think this approach aligns well with our main goal: 
we'd like the model to learn the underlying data distribution than memorize the correlation between the noise in the input
and the corresponding output.
Also, the reconstruction step can provide opportunities for new types of visualization, which we'll discuss more in section~\ref{sec:vis_latent_space}.

% branch off from each attribute embedding -> learn how to perform a 2D projection that can decode back to the attribute
We achieve this by defining an \emph{autoencoder}\cite{hinton2006reducing} for each attribute query $a_i$.
More specifically, we learn a projection to 2D through a series of layers, starting from the input layer, going to 2D,
denoted as an \emph{encoder} by $\mb{e}_i : \mathbb{R}^{d_i} \rightarrow \mathbb{R}^2$.
We also want to project back: reconstruct the original query (via its binary representation) from its 2D position, or a \emph{decoder}
$\mb{d}_i : \mathbb{R}^2 \rightarrow \mathbb{R}^{m(a_i)}$.
Then the first few layers in the encoder will be shared with the regressor as shown in Fig.~\ref{fig:nn_architecture}.

\textbf{Autoencoder Loss.} Since we represent attribute selections as binary-valued, a suitable loss function for measuring the quality of our
autoencoder is the binary cross-entropy loss:
\begin{equation}
	L_{ae} = -\sum_{j=1}^{m(a_i)} \left( \mb{r}(a_i)_j \log(\mb{d}_i(\mb{e_i}(z_i))_j) + (1-\mb{r}(a_i)_j) \log(1-\mb{d}_i(\mb{e_i}(z_i))_j) \right),
\end{equation}
where $z_i = f_i(\mb{r}(a_i))$ is the feature embedding of the query selection.
Note that this loss is defined for each attribute, in order to learn attribute-specific autoencoders.

\subsubsection{Combining Together}

We combine the prediction loss and the autoencoder loss to learn a function that can \emph{both} predict queries as well as learn 2D projections of attributes:
\begin{equation}
	L = L_{pred} + \lambda_3 L_{ae},
\end{equation}
where $\lambda_3$ is a weight giving importance to the autoencoder, relative to the weights on the prediction loss.
One can view this objective as a type of multi-task autoencoder\cite{ghifary2015domain}: we want to learn an embedding, and a 2D projection, that enables self-reconstruction, while
simultaneously learning to predict query aggregations.
Importantly, this permits us to \emph{contextualize} attribute selections with respect to the aggregation task.
The prediction task can be viewed as a form of supervision for the 2D projection task, thus attribute selections that result in similar predictions will have
similar feature embeddings, as well as similar 2D projections.

\section{Using \oursystem for Visual Exploration}
\label{sec:nnvis}

In this section, we describe how \oursystem can be used to build interactive data visualization systems.

\begin{figure}[!t]
	\centering
	\includegraphics[width=0.9\linewidth]{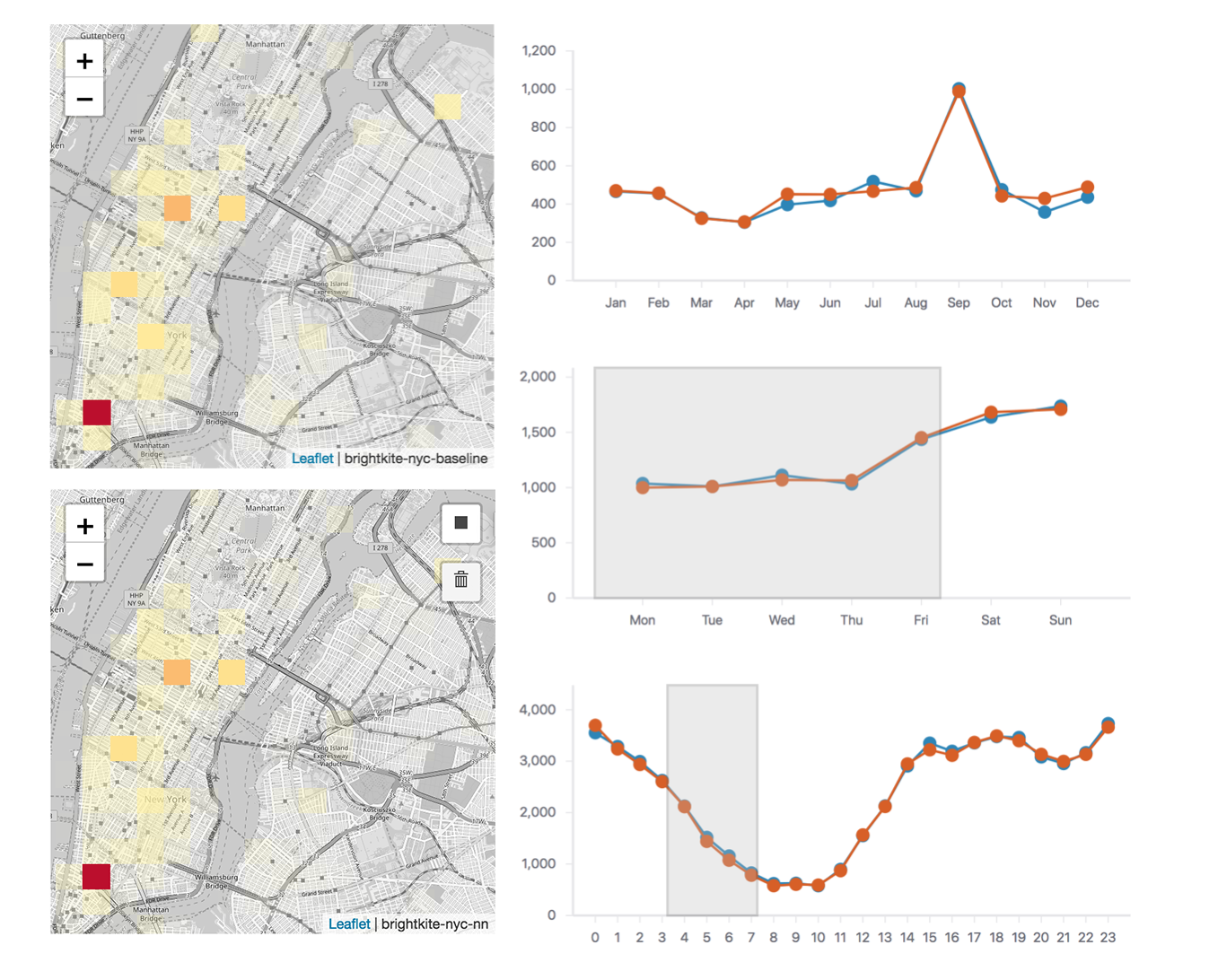}
  \caption{\oursystem can be used in a similar fashion to traditional data cubes techniques, allowing us to plot histograms and heatmaps with respect
	to various attribute selections.}
	\label{fig:nyc}
\end{figure}

\subsection{Plotting Histograms and Heatmaps with \oursystem}
\label{subsec:plot}

In traditional data cubes techniques, queries are typically made in order to plot histograms (for 1D attributes) and heatmaps (for 2D attributes).
This is typically realized through \textsf{group by} queries, where selections are made for all but one attribute, and then for the held-out
attribute, a single query is made to gather aggregations for each of its values, i.e.
\begin{verbatim}
SELECT COUNT(*) FROM BrightkiteTable 
GROUP_BY dayofweek
\end{verbatim}
\oursystem can enable the same type of visual exploration.
More specifically, we perform a \textsf{group by} query through our many-hot input encoding, placing a 1 on the attribute value that we would like to query,
and a 0 for all other attribute values.
Furthermore, we can take advantage of GPU data-parallelism in neural network implementations, and perform this operation in a single mini-batch,
providing a significant speed-up through GPU acceleration.
Our interface allows the user to perform arbitrary range selections for a given attribute, and enables interactive updates of histograms/heatmaps
over the remaining attributes, see Fig.~\ref{fig:nyc} for an illustration.

\subsection{Evaluating at Client Side}
Another advantage of \oursystem is that the trained model is small enough to be sent to client side for evaluation.
In comparison, other OLAP datacubes based techniques requires network connections with a backend server for interaction.
This advantage of \oursystem can be beneficial to both system users and service providers.
First, users can expect better experience when making queries.
Being able to evaluate at client side, \oursystem can eliminate network latency, which is usually a bottleneck.
Secondly, service providers can expect much lower cost because the same server can provide service to much more users since Query Per Second (QPS) will be significantly lower than other client-server systems.

\begin{figure}[!t]
	\centering
  \includegraphics[width=1\linewidth]{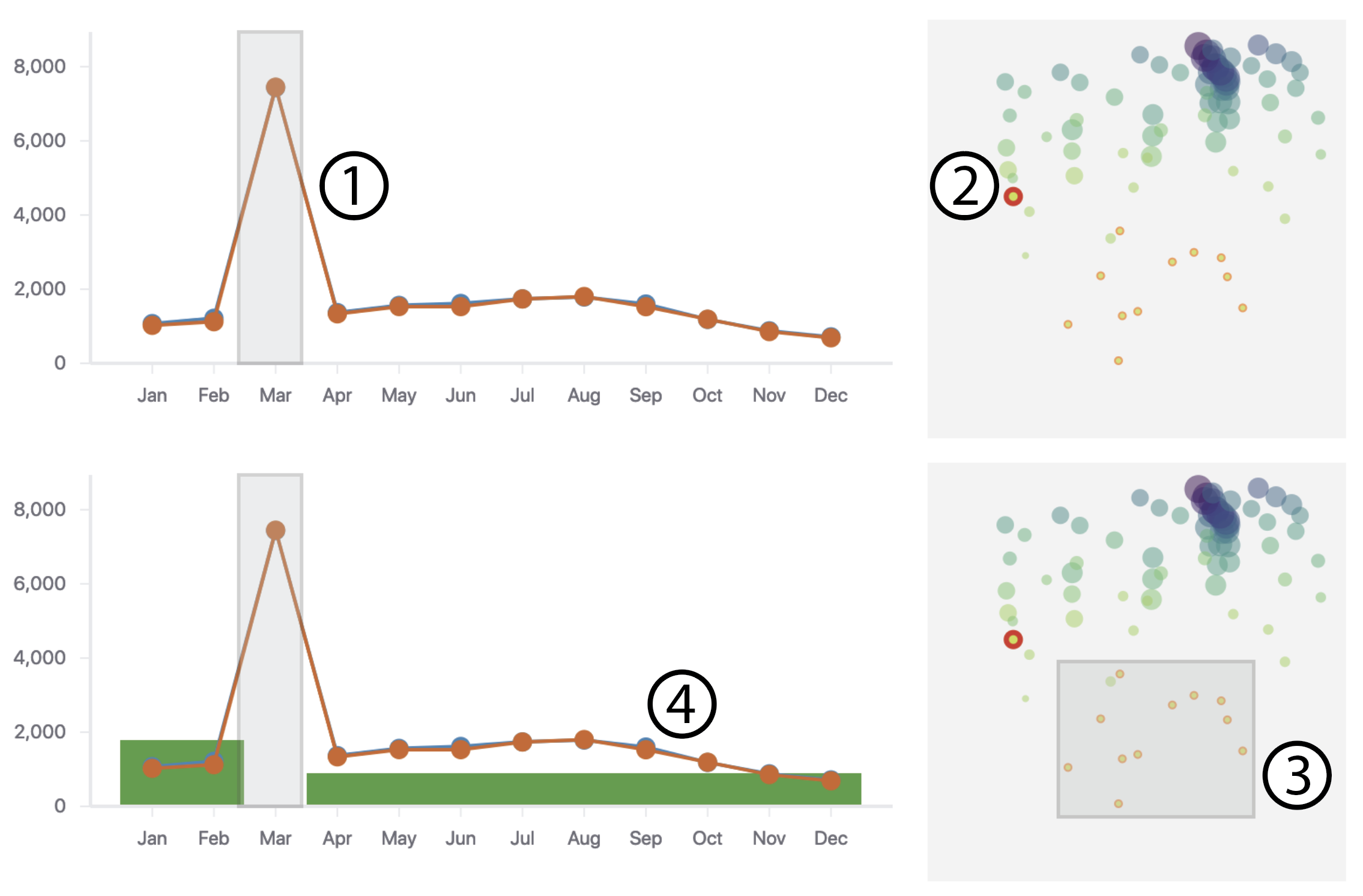}
  \caption{We show how the user can jointly interact with traditional histogram visualizations and latent space visualizations learned by \oursystem.
  On the left we initially show an histogram of counts of social media check-ins dataset of Austin for attribute \textit{Month}.
  Upon selecting month \textit{March} (1), the latent space on the right updates,
  to reflect this selection (2) (colored by red circular stroke with larger width) and 
  also highlight (colored by red circular stroke with smaller width) all possible selections with the same length of range (which is $1$ in this case). 
  By performing a selection in the latent space (3), 
  the user can explore the frequency of attribute selections (4) that belong to the selected latent space subset.}
  \label{fig:interaction}
\end{figure}

\subsection{Visualizing Attribute Latent Spaces}
\label{sec:vis_latent_space}

Although the network can replicate the types of queries perform with data cubes techniques, we can also use different structures
that the network learned to enable new forms of visual exploration.
In particular, we allow the user to explore the space of attribute selections through each attribute's learned 2D projection,
as discussed in Section~\ref{subsubsec:proj}.
To enable this, we first generate all possible ranges selections for a given attribute, and use the autoencoder
to create an overview of their distribution in the 2D space, where we visually encode attributes in a scatterplot.
Each point in the latent space view represent a selection of this attribute, where the radius of the point
is proportional to its aggregation value, and the color of the point represents the range of the selection, namely
the number of values selected in the attribute.
Importantly, we ensure that the latent space and the histograms/heatmaps discussed previously in Sec.~\ref{subsec:plot}
are \emph{linked}, so that interactions in one view update the other view, see Fig.~\ref{fig:interaction} for an example.

\section{Implementation}

\paragraph*{Software} \oursystem is written in Python 3.6 with PyTorch version 0.4.
It is implemented in the way that the neural network architecture is dynamically created given a JSON configuration file.
When changing a dataset or updating the neural networks, users only need to provide a new JSON configuration.
For a given trained model, we utilize it in a Flask http backend server, providing RESTful web services.
To test the functionality of evaluating at client side, we converted the model from PyTorch format to TensorFlow format.
Then we use TensorFlow.js in client side for model evaluation.
We have implemented a web user interface, \oursystem Viewer, for interaction, implemented in Javascript using React and React-Vis.
In order to make the training set generation faster, we performed ground truth aggregation queries in C++.

\paragraph*{Hardware} All the models are trained on a machine with an 8 core Intel i7-7700K 4.20GHz CPU, 32GB main memory, and a Nvidia GTX 1080 Ti GPU with 12GB video memory.

\paragraph*{Datasets} A summary of datasets and training/testing statistics of all the case studies is provided in Fig.~\ref{tb:experiments_summary}.
We evaluated our method on held-out test datasets, generated in the same manner as training data.
Testing error is computed as the average L1 norm difference between the predictions and ground truth, scaled by the inverse of the mean
of the ground truth set, in order to be commensurable across datasets.

\paragraph*{Architectures} While the architectures we used to train the models were all quite similar, there are some distinctions between them that we describe here for the sake of reproducibility. The table in Fig.~\ref{tb:architectures} describes the architectures used in each of the trained models we discuss in this paper.

\paragraph*{Mini-Batch}
When training neural networks, a mini-batch usually is constructed by randomly choosing a certain number of training samples.
However, in practice, we found training is not very stable using this approach for our setup.
One of the possible reasons could be the fact that the type of queries are unbalanced in the training set.
For example, for an attribute with 10 dimensions, there will be 10 queries needed to plot a line chart for it.
For an attribute with 5 dimensions, there will only be 5 queries.
Since our training set is generated by selecting the same number of random ranges for each attribute,
the number of queries for different attributes will be different.
Specifically, attributes with higher dimensionality will have more sample queries than attributes with lower dimensionality.
People have proposed solutions like stratified sampling\cite{zhao2014accelerating} to solve this problem.
In this paper, we take advantage of being able to control how we generate the training set and propose a \textit{state} based mini-batch construction strategy.
At training time, a mini-batch will be composed by the corresponding queries of several randomly selected \textit{states}.

\paragraph*{Training}
Although we need to tune parameters for each different model,
there are some empirical best practice that we found stay stable in our experiments.
When choosing the right weights $\lambda_1, \lambda_2, \lambda_3$, 
it usually works when setting $\lambda_1$ to be $1$. 
Next, set $\lambda_2$ so that $\lambda_2 L_1$ is one or two order of magnitude larger than $\lambda_1 L_{ae}$.
Then set $\lambda_3$ so that $\lambda_3 L_2$ is at least two order of magnitude smaller than $\lambda_2 L_1$.
Sometimes setting $\lambda_3$ to be $0.0$ achieves better performance.
When choosing the optimizer for training, we found if we only use mini-batch gradient descent(GD), 
it's hard for \oursystem to converge.
To solve this problem, we used a similar approach as described in\cite{keskar2017improving}, 
which simply use Adam\cite{kingma2014adam} for the first 10 to 20 epochs of training then switch to mini-batch gradient descent. 
This technique works well for all the training in our experiments.

\begin{figure}[t]
  \centering

  \begin{tabular}{|l|l|l|l|l|l|l|l|l|}\hline
    DataSet & Raw Data Size & \# States&  M. Size & RAE \\\hline
    B.K. NYC & 79k (3.1MB)    & 10k & 703KB & 4.25\% \\
    \hline
    B.K Austin & 22k (0.8MB) & 10k & 703KB & 3.88\% \\
    \hline
    Flights\_Count & 5m (204MB)& 60k & 1.2MB & 3.11\% \\
    \hline
    Flights\_Delay & 5m (204MB)& 60k & 1.2MB & 6.58\% \\
    \hline
    Yellow Cab & 12m (1.8GB)   & 30k  & 798KB & 0.97\% \\
    \hline
    SPLOM & 100k (3.9MB) & 10k &  135KB & 2.64\% \\
    \hline
  \end{tabular}
  \caption{Summary of training results for the experiments reported in this paper. 
    \emph{\# States} means the number of \emph{states} used in the training set.
    \emph{M. Size} represents the saved file size of trained models.
    \emph{RAE} represents Relative Absolute Error, which is calculated as
    $RAE$ = $\frac{\sum_{i} |\hat{y_i}-y_i|}{\sum_{i} |y_i-\bar{y}|} \times 100\%$,
    where $\hat{y_i}$ is the model's prediction for query $i$, $y_i$ is the ground truth for query $i$, 
    and $\bar{y}$ is the average value of $y_i$ in the testing set.}
  \label{tb:experiments_summary}
\end{figure}

\begin{figure}[t]
  \centering
  \begin{tabular}{|l|l|c|}
    \hline
    Input & Autoencoder & \multicolumn{1}{l|}{Regressor} \\ \hline
  \textbf{BrightKite} & & \\ \hline
    Month (12)      & {[}8, \textbf{4}, 2, 4,  8{]}      & \multirow{4}{*}{{[}220{]}}    \\ \cline{1-2}
    Day of Week (7) & {[}8, \textbf{4}, 2, 4, 8{]}       &                               \\ \cline{1-2}
    Hour (24)       & {[}12, \textbf{6}, 2, 6, 12{]}     &                               \\ \cline{1-2}
    Geospatial (40) & {[}400, \textbf{128}, 2, 128, 400{]} &                               \\ \hline
    \textbf{Flights (count)} & & \\
    \hline 
  Month (12)      & {[}120, \textbf{20}, 2, 20, 120{]}     & \multirow{6}{*}{{[}256, 128{]}}    \\ \cline{1-2}
  Day of Week (7) & {[}70, \textbf{20}, 2, 20, 70{]}     &                               \\ \cline{1-2}
  Hour (24)       & {[}240, \textbf{20}, 2, 20, 240{]}   &                               \\ \cline{1-2}
  Geospatial (40) & {[}400, 128, \textbf{20}, 2, 20, 128, 400{]} &                               \\ \cline{1-2}
  Carrier (10)    & {[}100, \textbf{20}, 2, 20, 100{]}     &                               \\ \cline{1-2}
  DelayBin (14)   & {[}140, \textbf{32}, 2, 32, 140{]}   &                               \\ \hline
    \textbf{Flights (delay)} & & \\
       \hline
       Month (12)      & {[}120, \textbf{20}, 2, 20, 120{]}     & \multirow{6}{*}{\makecell{{[}256, 128,\\ 64{]}}}    \\ \cline{1-2}
  Day of Week (7) & {[}70, \textbf{20}, 2, 20, 70{]}     &                               \\ \cline{1-2}
  Hour (24)       & {[}240, \textbf{20}, 2, 20, 240{]}   &                               \\ \cline{1-2}
  Geospatial (40) & {[}400, 128, \textbf{20}, 2, 20, 128, 400{]} &                               \\ \cline{1-2}
  Carrier (10)    & {[}100, \textbf{20}, 2, 20, 100{]}     &                               \\ \cline{1-2}
  DelayBin (14)   & {[}140, \textbf{32}, 2, 32, 140{]}   &                               \\ \hline
  \textbf{Yellow Cab} & & \\ \hline
    Month (12)      & {[}120, \textbf{20}, 2, 20,  120{]}      & \multirow{4}{*}{{[}220{]}}  \\ \cline{1-2}
    Day of Week (7) & {[}70, \textbf{20}, 2, 20, 70{]}       &                             \\ \cline{1-2}
    Hour (24)       & {[}240, \textbf{20}, 2, 20, 240{]}     &                             \\ \cline{1-2}
    Geospatial (40) & {[}400, 128, \textbf{20}, 2, 20, 128, 400{]} &                             \\ \hline
    \textbf{SPLOM} & & \\ \hline
    a0 (\#bin) & {[}16, \textbf{8}, 2, 8, 16{]} & \multirow{5}{*}{{[}120, 60{]}}    \\ \cline{1-2}
    a1 (\#bin) & {[}16, \textbf{8}, 2, 8, 16{]} &                               \\ \cline{1-2}
    a2 (\#bin) & {[}16, \textbf{8}, 2, 8, 16{]} &                               \\ \cline{1-2}
    a3 (\#bin) & {[}16, \textbf{8}, 2, 8, 16{]} &                               \\ \cline{1-2}
    a3 (\#bin) & {[}16, \textbf{8}, 2, 8, 16{]} &                               \\ \hline
  \end{tabular}

  \caption{Architectures for the neural networks used in the experiments with \oursystem.
    The numbers in bold represent attribute embedding layers as described in Fig.~\ref{fig:nn_architecture}.
  }
\label{tb:architectures}
\end{figure}

\section{Experimental Evaluation}
\label{sec:case_studies}

We have explored our method in the context of a variety of datasets, where in this section we highlight quantitative results of our method
in predicting aggregations, and qualitative results of our proposed visualization techniques.
We note through all results, we show our predictions as orange curves, and the ground truth aggregations in blue curves, in order
to consistently compare our method with the actual database queries.

\begin{figure*}[t]
	\centering
	\includegraphics[width=0.9\linewidth]{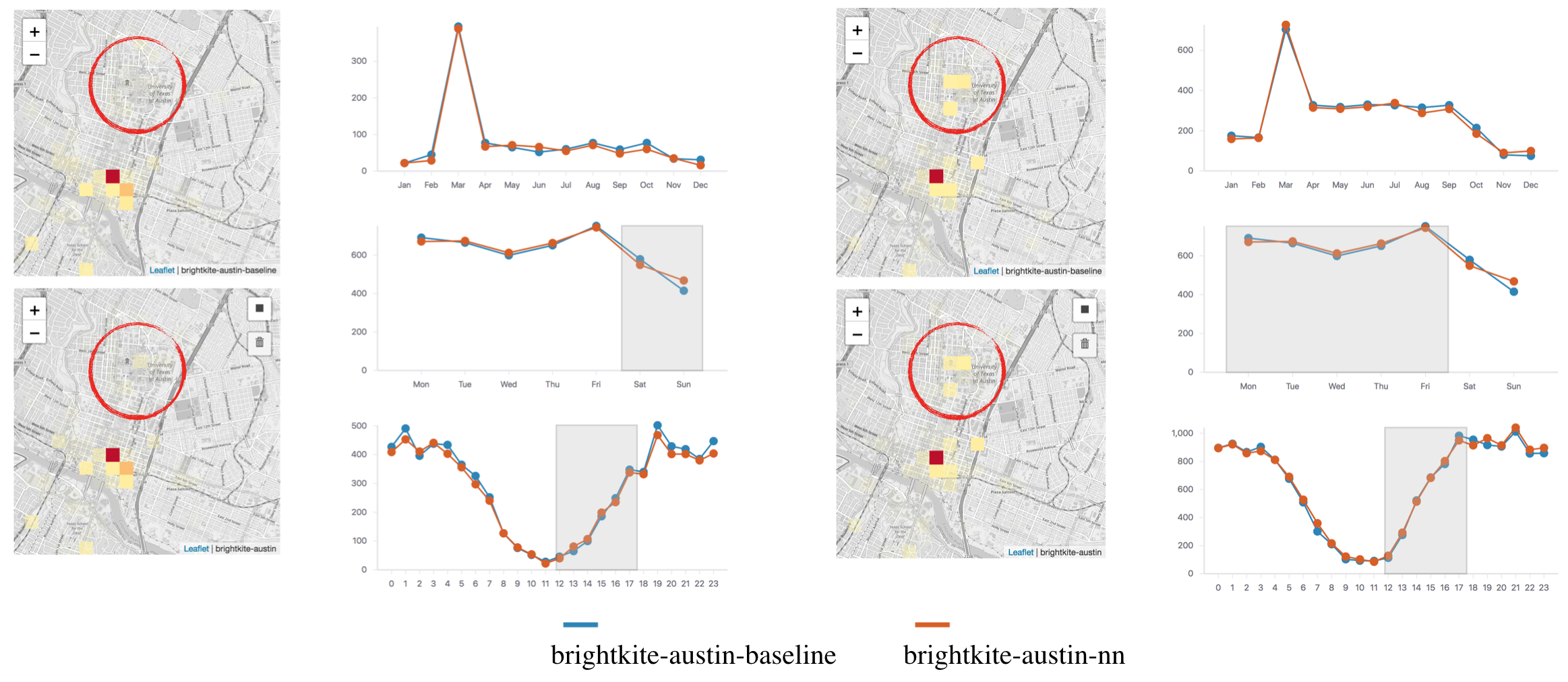}
  \caption{We show how \oursystem learned that there are more check-ins on University of Texas at Austin campus during daytime on weekdays.}
	\label{fig:utexas_campus}
\end{figure*}

\subsection{Brightkite Social Media Check-ins}

Here we study Brightkite\cite{cho2011friendship} social media check-ins to assess the
capability of \oursystem to learn the \textsf{count} aggregation. Fig.~\ref{fig:utexas_campus} shows
examples of how our technique can be treated analogous to a data cubes system, capable of
plotting histograms and heatmaps from aggregation queries.
We also use this dataset to evaluate training stability and to test the learned latent space contains meaningful information or not.

\subsubsection{Training Details} 
We choose two separate metropolitan areas and train separate networks on each: New York City and Austin.
We choose month, day of week, hour, and geospatial information (longitude and latitude) as input dimensions, following common practice in the study of urban activity\cite{lins2013nanocubes,miranda2017urban}.
The longitude and latitude are encoded as $20+20=40$ bins, following the strategy described in Fig.~\ref{fig:input_encoding}.
The weights for L1 loss and L2 loss for regressor and BCE loss for autoencoder are 20, 0.001, and 1 respectively.
Each model is trained for 1000 epochs and each epoch takes 6 seconds to train.

\subsubsection{Results} 
The predictions are good for both cities under the same neural network configuration.
This shows the generalizability of \oursystem.

\begin{figure}[h]
	\centering
	\includegraphics[width=\linewidth]{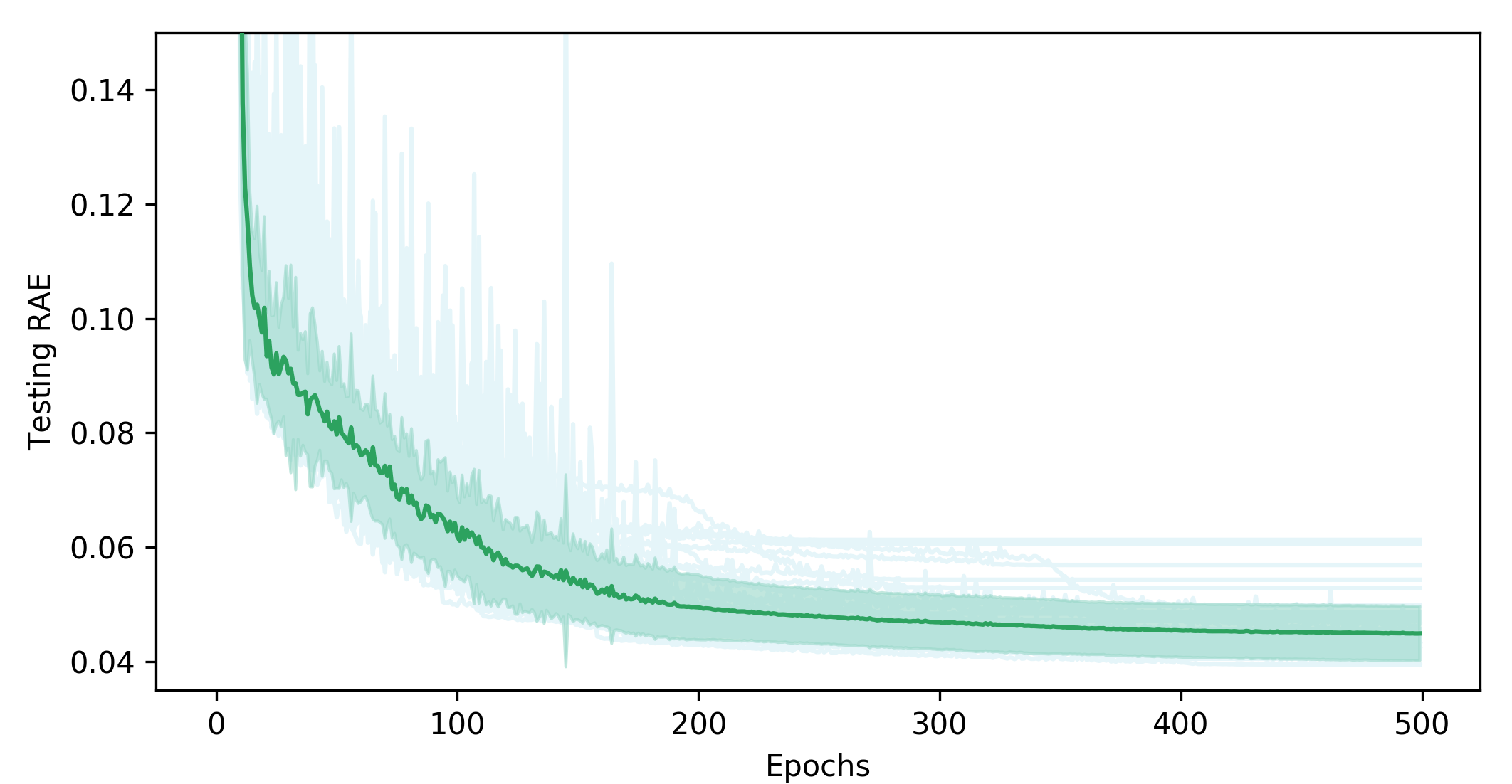}
  \caption{Testing error for 50 independent training on BrightKite NYC dataset with the same configuration.
    Light green lines represent each individual training.
    Solid green line represent the mean testing error.
    Darker green area represent standard deviation from mean.
    We can notice some variance between different runs. However, all of them eventually converge to relatively the same optimal.
  }
	\label{fig:multiple_training}
\end{figure}

\paragraph*{Training Stability} We also evaluated the training stability on BrightKite NYC dataset.
Specifically, we independently run 50 training on BrightKite NYC dataset with exactly the same configurations.
Then we record the testing error for each model for each epoch.
The results are shown in Fig.~\ref{fig:multiple_training}.
We can see that \oursystem always converge to the same optimal.

\begin{figure}[t]
	\centering
	\includegraphics[width=0.9\linewidth]{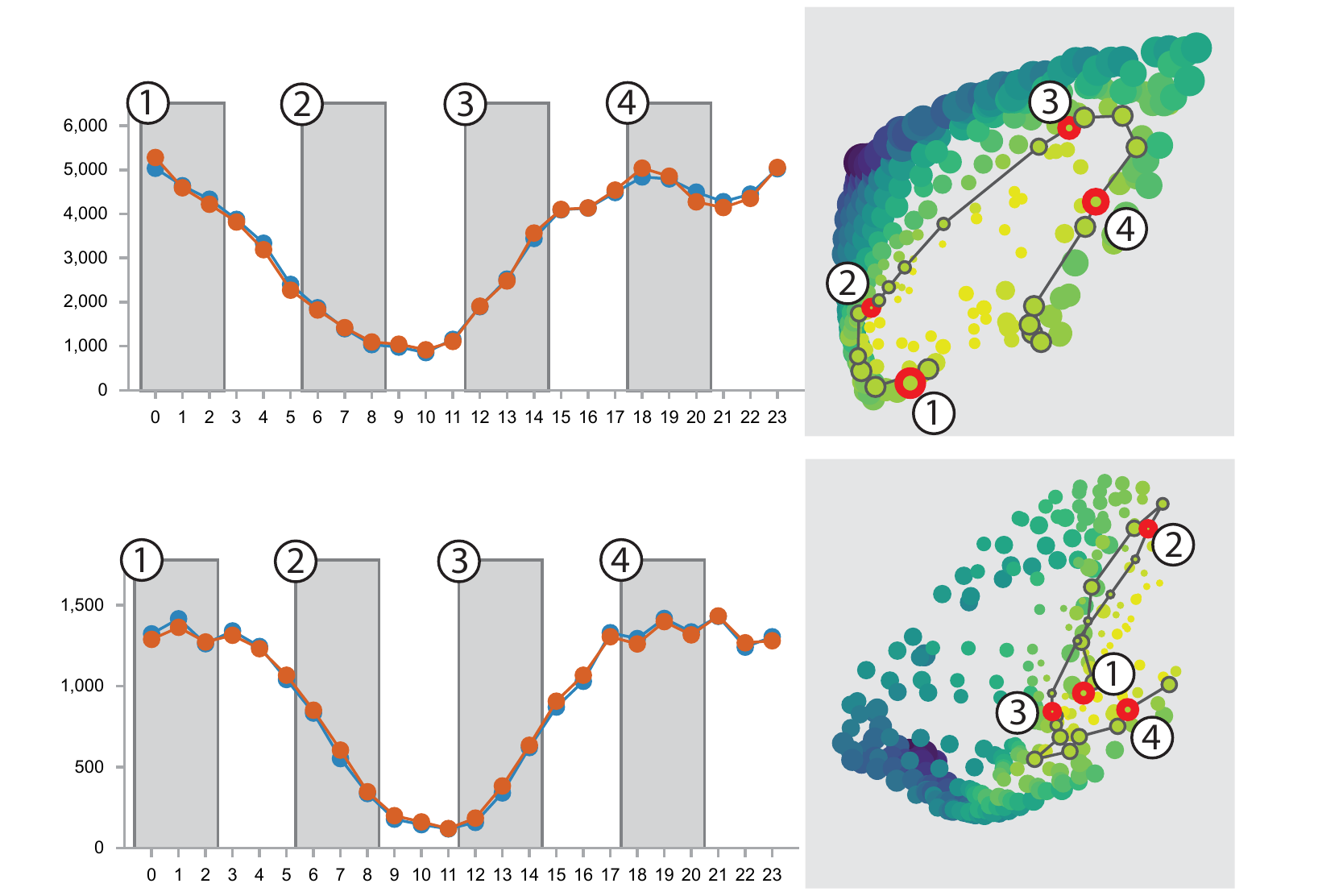}
  \caption{In this figure, we compare the diurnal patterns present in
	the latent space for ``hour'' in Austin and NYC. The top
	half shows the queries and latent space for NYC, while the
	bottom half shows them for Austin. We highlight the sequence
	of queries in the latent space for sliding ranges throughout
	the day, and notice that the activity in New York City is more
  circular in nature. We find this pattern always exists within multiple independently trained models; 
  though they may have different views (e.g. rotated, stretched) of the latent space due to randomness in the training.}
	\label{fig:compare_city}
\end{figure}

\paragraph*{Comparing Two Cities by Latent Space} 
The latent space plots may convey information that could not be directly perceived on histograms of counts.
Fig.~\ref{fig:compare_city} shows an example.
Firstly, the latent space for \emph{hour} of both cities forms a loop for ranges with same length.
This suggests that \oursystem learned the fact that \emph{hour} indeed has a repeated circular pattern.
However, the ``circle'' in Austin's latent space has a large opening.
This could be caused by the difference of lifestyles of the two cities: New York City never sleeps, while Austin goes to bed at night.

\subsection{Flight Dataset}
\label{sec:flights} 

We use a dataset collected by the Bureau of Transportation Statistics consisting of flight delay information in the year 2008\cite{flightdata}.
For this dataset, our first experiment uses the total flight counts as
the aggregation operation. Since this dataset contains an
attribute \emph{delay time}, which is also a meaningful attribute in
which to aggregate,
our second experiment builds a model to predict the average delay
time. Our goal with this model is to check the extent to which \oursystem can learn non-monotonic aggregations.

\subsubsection{Training Details}

\paragraph*{Count Predictions} For the \textsf{count} aggregation we filter flights to be within the contiguous United States, and restrict entries
to only the 10 most used airlines in the dataset, giving us a total of 5,092,321 entries after removing entries containing missing data.
We note that this dataset has more entries and attributes than the Brightkite dataset, including a numeric variable (Delay Time).
To encode the numeric variable in our many-hot encoding, we bin the delays in 15 minute increments.
The weights for L1 loss, L2 loss and BCE loss are 1, 1e-7, and 1, respectively.
Each model is trained for 500 epochs and each epoch takes 160 seconds to train.

\paragraph*{Average Predictions} We follow a similar training setup to \textsf{count}, with a couple exceptions.
Since generating training samples to predict average delay time itself is very time consuming,
we dropped longitude and latitude columns in the raw data and discarded entries whose delay time is smaller than $-60$ minutes or larger than $140$ minutes.
The weights for L1 loss, L2 loss and BCE loss are 10, 10, and 1, respectively.
Each model is trained for 500 epochs and each epoch takes 160 seconds to train.

\subsubsection{Results}

Fig.~\ref{tb:experiments_summary} shows the quantitative results for the two types of aggregation queries.
Overall, we find the errors to be competitive, if not lower, than Brightkite, showing that our method is capable
of handling different types of attributes, as well as different forms of aggregation.
Furthermore, for this dataset note that the size is quite large -- 204MB, while
the size of our trained networks only occupy 1.2MB for both tasks, demonstrating the significant compression capabilities of our network.

\subsection{YellowCab Taxi Dataset}

We use NYC YellowCab Taxi trip records of year 2015 from NYC Taxi and Limousine Commission (TLC)\cite{yellowcabdata} to study
the learning capacity of \oursystem in a series of controlled settings.

\subsubsection{Training Details}
We choose month, day of week, hour, and pickup location (longitude and latitude) as input dimensions.
The longitude and latitude are encoded as $20+20=40$ bins.
We created four different datasets under this same schema by sampling 1k records per month, 10k records per month,
100k records per month and 1 million records per month respectively from the original dataset.
We refer to these four datasets as YC-1K, YC-10K, YC-100K, and YC-1M respectively.
(We performed data cleaning and filtering after sampling. So the actual number of records of each month of a dataset will be slightly less than the sampled number.)
As described in Fig.~\ref{tb:architectures}, we use the same network configuration for the four datasets.
The weights for L1 loss, L2 loss and autoencoder loss are 100.0, 0.0 and 1.0 for YC-1K, YC-10K and YC-100K.
For YC-1M, the weights for L1 loss, L2 loss and autoencoder loss are 10.0, 0.0 and 1.0.
Each model is trained for 1000 epochs and each epoch takes 15 seconds to train.

\begin{figure}[t]
  \centering
  \begin{tabular}{|l|l|l|l|} \hline
    Raw Data Size & \# States & Model Size  & Testing RAE \\ \hline
    12k (1.8MB)   & 30k      & 798KB & 3.70\% \\
    120k (18MB)   & 30k      & 798KB & 2.04\% \\
    1.2m (180MB)  & 30k      & 798KB & 1.35\% \\
    12m (1.8GB)   & 30k      & 798KB & 0.97\% \\ \hline
  \end{tabular}
  \caption{Yellowcab dataset with different raw data size.}
  \label{tb:yc_raw_data_size}
\end{figure}

\begin{figure}[t]
  \centering
  \begin{tabular}{|l|l|l|} \hline
    Model Size & Training RAE & Testing RAE \\ \hline
       113KB        & 5.06\%       & 5.18\%      \\
       220KB        & 3.93\%       & 4.03\%      \\
       798KB        & 3.59\%       & 3.70\%      \\
       1.7MB        & 2.98\%       & 3.10\%      \\ \hline
  \end{tabular}
  \caption{Performance evaluation of different model size (e.g. different number of hidden layers) on the same dataset. In this experiment, we tested four different models on the Yellow Cab 12k dataset.}
  \label{tb:yc_model_size}
\end{figure}

\begin{figure}[t]
  \centering
  \begin{tabular}{|l|l|l|} \hline
    Training States & Training RAE & Testing RAE \\ \hline
       15k        & 6.74\% & 6.85\% \\
       30k        & 3.59\% & 3.70\% \\
       60k        & 3.39\% & 3.47\% \\ \hline
  \end{tabular}
  \caption{Performance evaluation of different training set size on the same dataset. In this experiment, we tested three different models on the Yellow Cab 12k dataset with different number of states in training sets.}
  \label{tb:yc_training_set_size}
\end{figure}

\subsubsection{Results}

\paragraph*{Raw Data Size} We training results for YC-1K, YC-10k, YC-100k and YC-1M are in Fig.~\ref{tb:yc_raw_data_size}.
We notice a fact that the testing error becomes smaller when the raw data size increase.
We think this is due to the fact that when more data are available, there will be less noise.
The data distributions that \oursystem need to learn will be more smooth, requiring less learning capacity.
Since we are using the same neural network configuration, we can expect lower error on dataset that doesn't require large learning capacity.

\paragraph*{Model Size} Another experiment we did is using neural networks with different sizes for the same dataset.
We start from a small neural network and then gradually increase the number of hidden layers and neurons to see if the prediction improves.
The results are in Fig.~\ref{tb:yc_model_size}.
Not surprisingly, with larger neural networks, the error reduces.

\paragraph*{Training Set Size} Another factor that can influence the training is the size of training set.
In this experiment, we tested the same model but with different number of training sets.
Results can be found in Fig.~\ref{tb:yc_training_set_size}.
We can see a significant accuracy improvement when using 30k training states than 15k training states.
However, the improvement from using 30k to 60k is very small.
The reason is that the capacity of the neural network may be not enough for 60k training states.

\subsection{SPLOM Dataset}

\begin{figure}[h]
  \centering
  \begin{tabular}{|l|l|l|l|} \hline
    Bin Size & \# States & Model Size  & Testing RAE \\ \hline
      10    &   10k     & 109KB       & 1.02\%      \\
      20    &   10k     & 116KB       & 1.85\%      \\
      30    &   10k     & 122KB       & 2.25\%      \\
      40    &   10k     & 129KB       & 2.09\%      \\
      50    &   10k     & 135KB       & 2.64\%      \\ \hline
  \end{tabular}
  \caption{SPLOM dataset with different bin size.}
  \label{tb:splom_experiments}
  \vspace{-1em}
\end{figure}

Last, we use the synthetic SPLOM dataset of Kandel et al.\cite{kandel2012profiler} to validate whether \oursystem can learn how to predict aggregational values
under a controlled setting.
Since all the attributes of SPLOM dataset are real values, it also provides us an opportunity to study the behavior of \oursystem when bin size increases.

\paragraph*{Training Details}
Following the procedure described in \cite{kandel2012profiler}, we generated $100,000$ entries of five-attribute records, and divide
each attribute into a prescribed number of bins.
We trained five different \oursystem using 10, 20, 30, 40, and 50 bins respectively.
The weights for L1 loss, L2 loss and autoencoder loss are 1.0, 0.0 and 1.0, respectively.

\paragraph*{Results}
A detailed quantitative evaluation of training \oursystem for SPLOM dataset can be found in the table of Fig.~\ref{tb:splom_experiments}.
As suggested in the table, when we increase the number of bins, the network requires more neurons, and thus a higher capacity, to learn well.
We note that the increase in testing error is to be expected, for several reasons.
The first reason is that when the bins are refined, few records fall into the same bin.
So the variance within each bin is larger, making it more difficult for \oursystem to learn the underlying distribution.
The second reason is that when the number of bins increase, the space of possible different queries grows exponentially.
Yet, we didn't increase the training set.

\section{Discussion}

The main limitation in \oursystem is the current dependence of the network architecture on the dataset and schema complexity.
While we were able to successfully train these networks with a certain amount of experimentation, the process is more cumbersome than we would like.
More automated methods to choose among different network architectures are still a current topic of research in machine learning, and beyond the scope of the current paper.

Another limitation is that the output of \oursystem are approximations. 
However, in a visual data exploration system, it is the overall trend and distribution that people mostly want to get.
Whenever the user needs the absolute correct answer, they can always query the database for that.
The value of \oursystem is that it provide a tool for them to quickly find what question they want to ask to the data.
Further more, \oursystem is designed for visualization.
The final results users get are heatmap, line charts, histograms, etc.
As long as the error is small, users can hardly tell the difference visually.

One notable feature of \oursystem\ is that the query training set is dependent on the affordances provided by the visual exploration system. This is an advantage in terms of machine learning, because the additional information available allows us to simplify the problem of training a network capable of answering \emph{any} query. At the same time, the fact that we have full control over the training set of queries is sometimes a disadvantage, because a poorly-generated training set can cause the training procedure to fail.
Our proof-of-concept system shows the advantages that a neural network provides in the context of a visual exploration system, but a thorough study on how to generate appropriate training sets remains a topic for future work.

\section{Conclusion and Future Work}

We believe that the main value of \oursystem\ lies in its ability to learn high-level features of the dataset from which powerful visual data exploration techniques become possible.
While the accuracy of the approximated results seems well within what is to be expected of a neural network, we warn against expecting that \oursystem\ would be able to learn minute details of the dataset. Liu et al.\cite{liu2013immens} distinguished between approximate, sampling-based systems and exact aggregation systems. We would consider \oursystem\ to be approximate as well, but approximate in its \emph{aggregations}; it trades exact accuracy of the answer for higher-level knowledge about the queries to be answered.

We remain enthusiastic about the future of connecting interactive information visualization and machine learning.
We are particularly interested in solving the above two problems by leveraging recent work in user modeling and predictive interaction\cite{battle2016dynamic}.
The better we can predict how people (through a particular UI) make queries against a DB in a visual exploration scenario, the more effectively we will be able to train \oursystem's networks.

Because the process of aggregating queries is a differentiable function, this opens up several opportunities, 
such as query sensitivity and the discovery of queries that lead to user-prescribed aggregations. 
We are excited to explore these research avenues as part of future work.

\section*{Acknowledgments}
%The icons used in diagrams are created by Noun Project users
%Knut M. Synstad, % (neural network icon)
%Creative Stall % (databse icon)
%and
%Muneer A.Safiah % (line plot)
%. (url: https://thenounproject.com/)

%The authors wish to thank A, B, and C. 
This work is supported in part by
the National Science Foundation (NSF) under grant numbers IIS-1452977, IIS-1513651 and IIS-1815238;
and by the Defense Advanced Research Projects Agency (DARPA) under agreement numbers FA8750-17-2-0107 and FA8750-19-C-0002;
%JAL's acknowledgements
and by the U.S. Department of Energy, Office of Science, Office of Advanced Scientific Computing Research, under Award Number(s) DE-SC-0019039.

%The authors wish to thank A, B, and C. This work was supported in part by
%a grant from XYZ (\# 12345-67890).
%%JAL's acknowledgements
%This material is based upon work supported by the U.S. Department of Energy, Office of Science, Office of Advanced Scientific Computing Research, under Award Number(s) DE-SC-0019039.}

%\bibliographystyle{abbrv}
\bibliographystyle{abbrv-doi}

\bibliography{ref}
\end{document}